\documentclass[10pt,conference]{IEEEtran}
\IEEEoverridecommandlockouts
\usepackage[T1]{fontenc}
\usepackage{graphicx,times,cite,amsmath, amssymb, epsfig, array, mathrsfs}
\usepackage{array}
\usepackage{textcomp}
\usepackage{subfig}
\usepackage{url}
\usepackage{verbatim}
\usepackage{multirow}                
\usepackage{amsmath}
\usepackage{xcolor}
\usepackage{wasysym} 
\usepackage[linesnumbered, ruled]{algorithm2e}  
\SetKwRepeat{Do}{do}{while}

\newtheorem{theory}{Theorem}
\newtheorem{lemma}{Lemma}

\newlength{\figwidth}
\begin{document}
	
	\title{
  Opportunistic Subarray Grouping for RIS-Aided Massive Random Access in Cellular  Connectivity
 \vspace{-2mm} }
\author{
Yizhu Wang\IEEEauthorrefmark{1},  
Zhou~Zhang\IEEEauthorrefmark{2}, 
Saman Atapattu\IEEEauthorrefmark{3}
and Marco~Di~Renzo\IEEEauthorrefmark{4}\\
 \IEEEauthorblockA{
\IEEEauthorrefmark{1}Tianjin Artificial Intelligence Innovation Center (TAIIC), Tianjin, China.  \\
 \IEEEauthorrefmark{2} Hunan University, Changsha, Hunan, China.\\
 \IEEEauthorrefmark{3}School of Engineering, RMIT University, Melbourne, Victoria, Australia.  \\
 \IEEEauthorrefmark{4}Université Paris-Saclay, 3 Rue Joliot Curie, 91190 Gif-sur-Yvette, France. \\
\IEEEauthorblockA{
\IEEEauthorrefmark{1} wangyizhuj@163.com;\,
\IEEEauthorrefmark{2} zt.sy1986@163.com;\,
\IEEEauthorrefmark{3}saman.atapattu@rmit.edu.au;\,
\IEEEauthorrefmark{4}marco.di-renzo@universite-paris-saclay.fr
}
}
\vspace{-0mm}
\vspace{-0.75cm} }

\maketitle
\begin{abstract}
In Reconfigurable Intelligent Surfaces (RIS), reflective elements (REs) are typically configured as a single array, but as RE numbers increase, this approach incurs high overhead for optimal configuration. {\it Subarray grouping} provides an effective tradeoff between performance and overhead. This paper studies RIS-aided massive random access (RA)  at the Medium Access Control (MAC) layer in cellular networks to enhance throughput. We introduce an opportunistic scheduling scheme that integrates multi-round access requests, subarray grouping for efficient RIS link acquisition, and multi-user data transmission. To optimize access request timing, RIS estimation overhead and throughput, we propose a multi-user RA strategy using sequential decision optimization to maximize average system throughput. A low-complexity algorithm is also developed for practical implementation. Both theoretical analysis and numerical simulations demonstrate that the proposed strategy significantly outperforms the extremes of {\it full-array grouping} and {\it element-wise grouping}. 
\end{abstract}
	

\vspace{-2mm}	
\section{Introduction}
\vspace{-1mm}	
Reconfigurable Intelligent Surfaces (RIS) are a transformative technology for enhancing throughput, spectral efficiency, and coverage in wireless networks. By controlling a large array of Reflecting Elements (REs), RIS dynamically optimizes signal propagation, enabling constructive and destructive interference for improved performance, energy efficiency, and coverage. While extensive research has focused on the PHY layer of RIS~\cite{Fang2022,Katsanos2022} and some on subarray grouping~\cite{Kundu2022}, there is limited work on RIS optimization at the MAC layer~\cite{CaoMar2021,Li2020,wangz2019}. In massive random access (RA) networks, such as Cellular IoT, the dynamic nature of user requests and varying channel conditions adds complexity and increases the overhead of RIS channel estimation. Efficient base station (BS) scheduling is essential to maximize RIS performance, yet remains largely unexplored. Developing an effective RIS-aided RA framework is key to optimizing network efficiency.
\vspace{-2mm}	
\subsection{Related Work}
\vspace{-1mm}
In massive RA networks, efficient multi-user access and resource allocation have been widely researched. To address massive RA in  machine-type communication networks, tagged preambles and non-orthogonal multiple access  have been used to maximize throughput \cite{Wang2020}, along with a collision resolution scheme for RA collisions in cellular networks \cite{Althumali2020}. In RIS-aided networks, joint beamforming has enhanced multi-user access, while CSMA-based TDMA and FDMA strategies maximize uplink throughput \cite{CaoMar2021}. Multi-RIS networks also improve BS communication with cell-edge users, optimizing weighted sum rates through joint BS beamforming and RIS phase shifts \cite{Li2020}, and a heuristic distributed method in multi-cell systems enhances multi-BS power allocation and RIS reflection for throughput maximization \cite{Katsanos2022}.
Moreover, RIS channel estimation overhead significantly affects system performance due to the high number of REs and lack of RF chains, making efficient CSI acquisition a major challenge. To address this, an RE on-off switching strategy was proposed in \cite{wangz2019}, activating one element at a time for RIS channel estimation. To reduce pilot overhead, \cite{Zheng2020TC} introduced a subarray grouping approach, dividing REs into groups for joint channel estimation and beamforming, minimizing pilot symbols with allowable performance loss. In \cite{Kundu2022}, a closed-form upper bound for the ergodic rate with RE groups was derived, demonstrating that optimal grouping outperforms individual RE estimation.  
\vspace{-2mm}	
\subsection{Problem Statement and Contributions}
\vspace{-1mm} 

Subarray grouping strikes a balance between performance and overhead by dividing the RIS into manageable subgroups. It reduces overhead compared to element-wise grouping while offering more flexibility than full array grouping, making it ideal for large-scale RIS deployments. Integrating subarray grouping for RIS-aided massive RA at the MAC layer in cellular networks presents several challenges:  
(i) {\it User Dynamics}: Sporadic access can cause congestion, straining network capacity;  
(ii) {\it RIS Channel Estimation Overhead}: High RE counts and limited RF chains make CSI acquisition time-consuming, impacting performance;  
(iii)  {\it Dynamic Channel Conditions}: Time-varying channels may favor direct user-BS links, complicating RIS-based transmission decisions.

This work proposes an opportunistic RIS-aided RA strategy to improve multi-user efficiency, reduce CSI overhead, and adapt to dynamic channels. Key contributions include:  
(i) Designing an overhead-aware, opportunistic massive RA framework with flexible user scheduling, subarray-group adaptation, and efficient RIS-aided transmission;  
(ii) Proposing an optimal RIS-aided RA strategy formulated as a sequential decision-making task to maximize system throughput using a two-level threshold structure;  and
(iii) Developing a low-complexity RA algorithm, including offline throughput calculation and an online RA method, with results validating its effectiveness.

  \vspace{-2mm}
\section{System model}\label{s:system_model}
\vspace{-2mm}
\subsection{Network Model}  
\vspace{-1mm}

\begin{figure}[t!]
	\begin{center}
		\includegraphics[scale=.35]{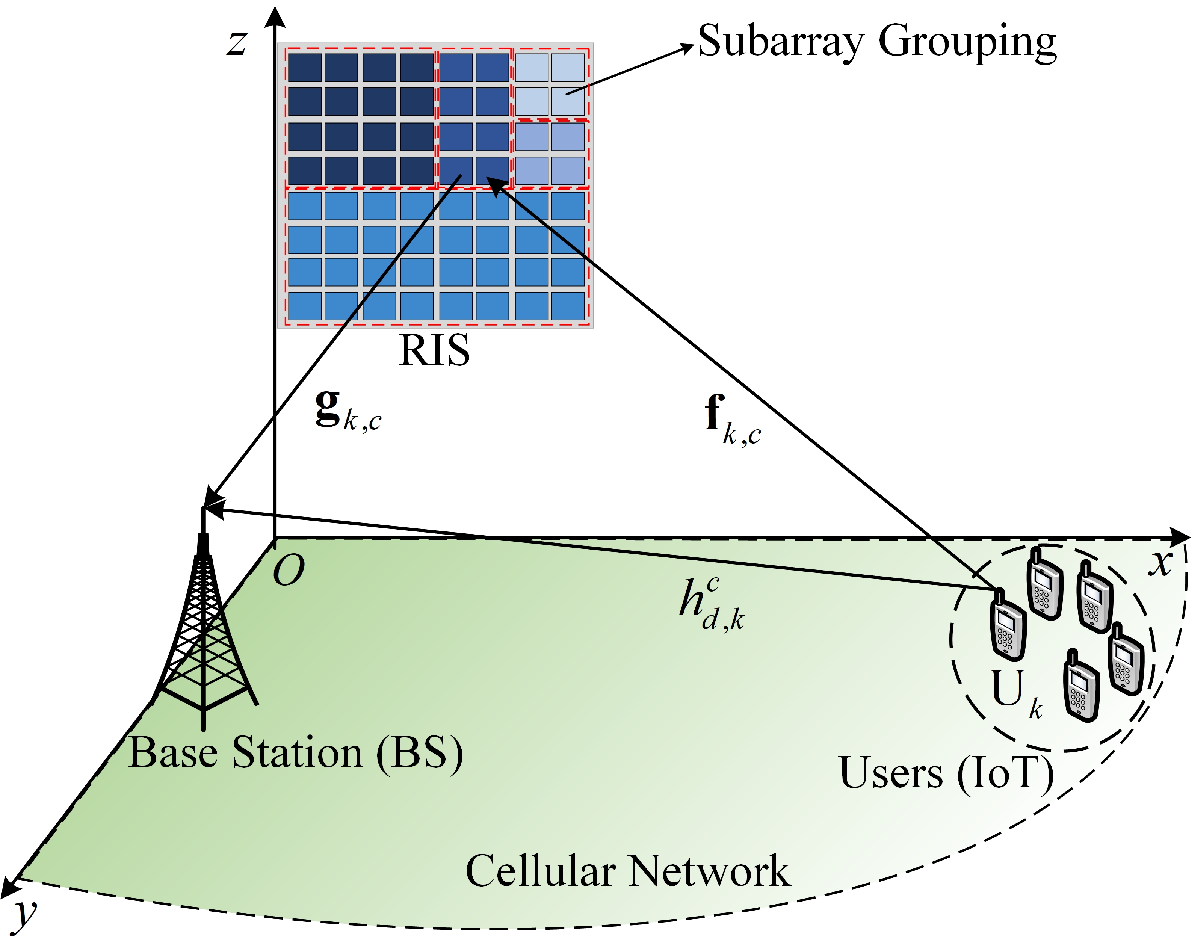}
		\caption{Subarray grouping RIS-aided cellular connectivity. 
        }
        \label{f:IRS_estimation}
	\end{center}
 \vspace{-3mm}
\end{figure}

We consider a wireless cellular network with \( K \) users and a BS, as shown in Fig.~\ref{f:IRS_estimation}. Each user, \(\text{U}_k\), \( k=1, \cdots, K \), accesses the BS via a rectangular RIS with 
 \( M = 2^{\overline{J}+1} \) REs, 
coordinated by a BS-connected controller, where $\overline{J}$ denotes the maximal subarray grouping levels.
The user-BS link distance \( d_s \), user-RIS link \( d_{r,1} \), and RIS-BS link \( d_{r,2} \) define the distances. Path loss exponents for direct and RIS links are \(\alpha_1\) and \(\alpha_2\), with reference path loss \(\beta_0\).
\addtolength{\topmargin}{0.02in}
For the wireless link model, we assume frequency-selective fading channels, with users transmitting data to the BS using orthogonal frequency division multiple access (OFDMA) and a transmit power \( P_t \). There are \( C \) sub-channels, and for each sub-channel \( c = 1, \cdots, C \), user-BS channel gains are \( h_{d,k}^c \); RIS link channel gains are \(\mathbf{f}_{k,c} = [{f}_{k,c}^m] \in \mathbb{C}^{M \times 1}\) and \(\mathbf{g}_{k,c} = [{g}_{k,c}^m] \in \mathbb{C}^{M \times 1}\), where \( m = 1, \ldots, M \) represents the \( m \)th RE of the RIS.
All links experience quasi-static Rayleigh fading, with gains following circularly symmetric complex Gaussian (CSCG) distributions: \( h_{d,k}^c \sim \mathcal{CN}(0, d_{s}^{-\alpha_1}) \), \( f_{k,c}^m \sim \mathcal{CN}(0, d_{r,1}^{-\alpha_2}) \), and \( g_{k,c}^m \sim \mathcal{CN}(0, d_{r,2}^{-\alpha_2}) \). Gains remain constant over the coherence time \( T_c \), and the noise power at the RIS and BS is \( N_0 \).

\begin{figure*}
	\begin{center}
		\includegraphics[scale=.75]{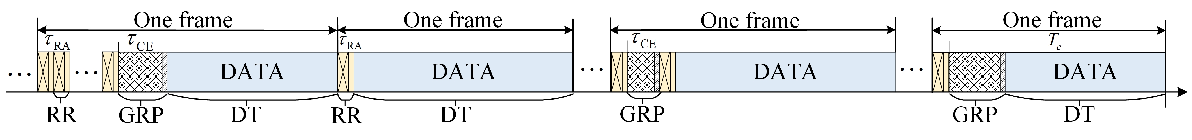}
		\caption{{Time diagram illustrating diverse phases cases across different time frames}}\label{f:super_frame}
	\end{center}
\vspace{-8mm}
\end{figure*}

\vspace{-2mm}

\subsection{Proposed Multi-Phase RA Scheme}\label{sub:proposed_scheme}

In this network, \( K \) users perform random access in a time-frame-based manner coordinated by the BS, as shown in Fig.~\ref{f:super_frame}. Each time frame represents a complete data transmission round from requested users to the BS. We propose a multi-phase RA scheme for each time frame, consisting of the following phases: {\it Random Request} (RR), {\it Grouping-based RIS Probing} (GRP), and {\it data transmission} (DT). These phases follow a sequential decision framework, detailed as follows.

At the start of each time frame, the \( K \) users randomly select one of \( S \) preambles with a probability of \( 1/S \) for access grant during the Random Request (RR) phase, lasting \( \tau_{\rm{RA}} \)~\cite{Althumali2020}. These preambles serve as initial user identifiers. If a preamble is selected by only one user, it results in a successful channel grant; however, if multiple users select the same preamble, a collision occurs, leading to grant failure. 

In the \( n \)th RR phase, the BS detects preambles from \( K_n \) users in set \( \mathcal{I}_n \) with expected size \( \mathbb{E}[K_n] = K \left((S-1)/S\right)^{K-1} \), and obtains user-BS instantaneous channel gains \( \mathbf{h}_{d}(n) = [\mathbf{h}_{d,1}(n), \dots, \mathbf{h}_{d,K_n}(n)] \), where \( \mathbf{h}_{d,k}(n) = [h_{d,k}^c(n)]\in\mathbb{C}^{C\times 1}\)  represents the \( k \)th user’s channel gain vector in \( \mathcal{I}_n \). The BS then has three first-layer decision options: 
\begin{enumerate}
    \item The BS may stop further phases and initiate the DT phase by scheduling users as \( \mathbf{a}_n = [a_{n,1}, \ldots, a_{n,c}] \) with \( a_{n,c} \in \mathcal{I}_n \cup \{0\} \). Here, \( a_{n,c} \in \mathcal{I}_n \) indicates sub-channel \( c \) is allocated to user \( a_{n,c} \), and \( a_{n,c} = 0 \) means sub-channel \( c \) is unused. Data transmission occurs without the RIS (detailed in Sec. \ref{sub:direct_multi_ch}). Based on \( \{{\mathcal I}_n, \mathbf{h}_{d}(n)\} \), the immediate reward is \( y_{n}({\mathcal I}_{n}, \mathbf{h}_{d}(n)) = T_c \cdot R_{d,n}^* \), where \( R_{d,n}^* \) is the achievable rate from (\ref{op:maximal_rated}); 
\item The BS may proceed to the \((n+1)\)th RR phase, aiming for a potentially higher future reward \( y_{n+1} \) than \( y_n \), accepting the cost of waiting for additional phases;
\item The BS may start the GRP phase to probe RIS links, then decide to either initiate the DT phase or proceed to the next RR phase. A subarray grouping scheme is applied to balance probing overhead and RIS transmission gain~\cite{Kundu2022}. The BS sets a probing vector \(\mathbf{b}_n = [b_{n,1},\ldots,b_{n,C}]\), where \(b_{n,c} \in \mathcal{I}_n\) means sub-channel \(c\) is probed by user \(b_{n,c}\), and \(b_{n,c} = 0\) indicates no probing. The BS also selects the subarray grouping level \(J_n \in [1,...,\overline{J}]\). 
In the GRP phase, the BS configures the RIS beamforming vector by grouping subarrays at level \(J_n\), with subarray containing \(B_n = 2^{\overline{J}-J_n}\) REs. The BS then instructs scheduled users to send pilots based on decision variable \(\mathbf{b}_n\) for channel estimation, spending \(\tau_{\rm{CE}}(J_n) = 2^{J_n+1} \tau_s\) for the process, where \(\tau_s\) is the pilot duration for one RE, as detailed in Sec. \ref{sub:csi_acq}. 
By collecting pilots, the BS obtains the grouping-based effective cascaded RIS channel gains, denoted as
{$\mathbf{h}_{r}^{J_n}(n)=[\mathbf{h}_{r,b_{n,c},c}^{J_n}(n)]_{c=1,...,C,b_{n,c}\neq 0}$, where
$\mathbf{h}_{r,b_{n,c},c}^{J_n}(n)=[{h}_{b_{n,c},c,u}^{J_n}(n)]\in\mathbb{C}^{2^{J_n}\times 1}
$}, and ${h}_{b_{n,c},c,u}^{J_n}(n)=\sum\limits_{m=1}^{B_n}
f_{b_{n,c},c}^{(u-1)B_n+m}(n)g_{b_{n,c},c}^{(u\!-\!1)B_n+m}(n)$. 
\end{enumerate}

After the GRP phase, the BS has two second-layer decision options:
\begin{enumerate}
    \item It may initiate the DT phase by scheduling users for RIS-aided transmission (described in Sec. \ref{sub:csi_acq}). The immediate reward for data transmission is $y_{n}\big({\mathcal I}_{n},\mathbf{h}_{d}(n),\mathbf{b}_n,\mathbf{h}_{r}^{J_n}(n)\big)\!=\!
(T_{c}\!-\!\tau_{\rm{CE}}(J_n))R_{r,n}(\mathbf{b}_n)$, where RIS-aided rate $R_{r,n}(\mathbf{b}_n)$ is derived in (\ref{equ:R_rate}); 
\item It may continue to the \((n+1)\)-th RR phase, hoping for a higher reward \( y_{n+1} \) at the cost of waiting.
\end{enumerate}
After one DT phase, this frame ends and next frame begins.

The proposed RA scheme enables multiple RR and GRP phases before the final DT phase, offering flexibility in utilizing user-BS and RIS links, which vary over time and users. By allowing the BS to wait for better conditions, higher throughput gains are achieved in the DT phase.


\vspace{-1mm}
\subsection{Adaptive Transmission Schemes} 
\vspace{-1mm}
After the $n$th RR phase within a time frame, the BS schedules users in set $\mathcal{I}_n$ to transmit data using OFDMA during the DT phase. Two transmission schemes are considered, depending on the use of RIS links.

\subsubsection{User-BS Transmission Scheme}\label{sub:direct_multi_ch}  
Based on user-BS link CSI $\mathbf{h}_{d}(n)$ for users in $\mathcal{I}_n$ acquired during the RR phase, the BS schedules users via the vector $\mathbf{a}_n$ for transmission without RIS. The instantaneous sum-rate $R_{d,n}(\mathbf{a}_n)$ is given by 
\vspace{-3mm}
\begin{equation}\label{op:maximal_rated}
\setlength\belowdisplayskip{3pt}
 R_{d,n}(\mathbf{a}_n)=  \sum_{c=1}^{C}\mathbf{I}[a_{n,c}\neq 0]\log_2\Big(1+\overline{\gamma}|h_{d,{a_{n,c}}}^c(n)|^2\Big)
\end{equation}
where $\overline{\gamma} = \frac{P_t \beta_0}{N_0}$, and the data transmission time is $T_c$.

To find optimal scheduling vector $\mathbf{a}_n^*$ maximizing sum-rate $R_{d,n}$, we need to solve a integer optimization problem with exponential complexity \cite{Yin2016}.
We adopt greedy algorithm in Algorithm \ref{Algorithm0} which provides low complexity and sub-optimal solution $\mathbf{a}_n^*$.
The algorithm utilizes $DesSort(\cdot)$, a descending-order sorting function, $Index(\cdot)$, sub-channel and user index retrieval function,
and $Find(\mathcal{C},c)$, a location function for an element.   
Correspondingly, the maximal sum-rate is expressed as $R_{d,n}^*=R_{d,n}(\mathbf{a}_n^*)$.  
 \vspace{-2mm}
 \begin{algorithm}
	\caption{{Greedy algorithm for $\mathbf{a}_n^*$}}\label{Algorithm0}
	\SetKwInOut{Input}{Input}\SetKwInOut{Output}{Output}
	\Input{${\mathcal I}_n,\mathbf{h}_{d}(n)$}

          ${\mathbf{h}}\leftarrow DesSort([\mathbf{h}_{d,1}^T,\ldots, \mathbf{h}_{d,K_n}^T]$)
          
          $\mathcal{K}\leftarrow [~]$, $\mathcal{C}\leftarrow [~]$, $i=1$, $\ell=1$, $\mathbf{a}_n^*=\mathbf{0}$
          
           \While{$\ell\le \min(K_n, C)$}{
           $(c^\dag,k^\dag)\leftarrow$ $Index({\mathbf{h}}(i))$
                 
                 \If{$c^\dag\notin \mathcal{C}~\&~k^\dag\notin \mathcal{K} $}        
                 {$\mathcal{C}\leftarrow [\mathcal{C},c^\dag]$, $\mathcal{K}\leftarrow [\mathcal{K},k^\dag]$,
                 $\ell\leftarrow \ell+1$
                 }           
           $i\leftarrow i+1$
           }
           $a_{n,c}^*\leftarrow \mathcal{K}(Find(\mathcal{C},c))$, $c=1,...,C$          
\end{algorithm}
\vspace{-2mm}
\subsubsection{RIS-aided transmission scheme}\label{sub:csi_acq}
For the subarray grouping based beamforming transmission, we denote reflection coefficient vector over sub-channel $c$ as $\mathbf{\Phi}_c=[\phi_{1,c},...,\phi_{M,c}]^T\in \mathbb{C}^{M\times 1}$,
where $\phi_{m,c}=\beta_{m,c} e^{j\theta_{m,c}}$ represents the reflection coefficient of $m$th RE, consisting of amplitude coefficient $\beta_{m,c}\in[0,1]$ and phase shift $\theta_{m,c}\in[-\pi,\pi)$, satisfying $|\phi_{m,c}|\le1, \forall m=1,...,M$.

Under subarray grouping level $J_n$, BS divides the RIS into $2^{J_n}$ subarrays of $B_n$ REs. 
In subarray $u=1,...,2^{J_n}$, REs are continuous, indexed from $(u-1)B_n+1$ to $u B_n$. 
As all REs in the same subarray use a common reflection coefficient,
the reflection coefficients $\mathbf{\Phi}_c$ over sub-channel $c$ can be expressed as $\mathbf{\Phi}_c=\overline{\mathbf{\Phi}}_c(J_n)\otimes\mathbf{1}_{B_n\times1}$,
where $\overline{\mathbf{\Phi}}_c(J_n)=[\overline{\phi}_{u,c}]_{u=1,...,2^{J_n}}$ denotes the effective
reflection coefficient vector with $\overline{\phi}_{u,c}$ representing the common refection coefficient for subarray $u$.

As the BS acquires the cascaded channel gains $\mathbf{h}_{r,b_{n,c},c}^{J_n}(n)$,
the combined channel gain by RIS beamforming vector $\mathbf{\Phi}_c$ over sub-channel $c$ from the probed user $b_{n,c}$ to the BS can be expressed as $\overline{\mathbf{\Phi}}_c^T(J_n) \cdot \mathbf{h}_{r,b_{n,c},c}^{J_n}(n)$.

Based on acquired RIS link CSI $\mathbf{h}_{r,k}^{J_n}(n)$ during the GRP phase, 
the BS finds the optimal beamforming vectors.

For sub-channel $c$, the optimal RIS beamforming vector $\overline{\mathbf{\Phi}}_c^*(J_n)$ maximizing the RIS-aided transmission rate 
can be found by solving the following optimization problem:
\vspace{-3mm}
\begin{align}\label{op:maximal_rate}
	& \max\limits_{\overline{\mathbf{\Phi}}_c} \Big|h_{d,b_{n,c}}^c(n)+\sum\limits_{u=1}^{2^{J_n}} {h}_{b_{n,c},c,u}^{J_n}(n) \overline{\phi}_{u,c}
 \Big|
	\\&
	\text{s.t.}~~ 0\le \overline{\theta}_{u,c}\le 2\pi,  0\le \overline{\beta}_{u,c}\le 1, u=1,...,2^{J_n}.\nonumber
\end{align}
The optimal value is $\overline{\mathbf{\Phi}}_c^*(J_n)=[\overline{\phi}_{1,c}^*,...,\overline{\phi}_{2^{J_n},c}^*]^T$,
where $\overline{\phi}_{u,c}^*=e^{j\overline{\theta}_{u}^*}$ with $\overline{\beta}_{u,c}^*=1$ and 
$\overline{\theta}_{u,c}^*= \text{mod} [\arg{h_{d,b_{n,c}}^c}-\arg{{h}_{b_{n,c},c,u}^{J_n}(n)},2\pi]$. 

Therefore, by probing users in vector $\mathbf{b}_n$,
the instantaneous sum-rate $R_{r,n}$ for RIS-aided transmission can be expressed as
\vspace{-6mm}
\begin{align}\label{equ:R_rate}
\setlength\belowdisplayskip{3pt}
\vspace{-4mm}
	&R_{r,n}(\mathbf{b}_n)=
 \sum\limits_{c=1}^C 
	 \Big(\mathbb{I}[b_{n,c}\neq 0]\nonumber\\&
  \vspace{-4mm}
  \quad\log_2\big(
	1+ 
 \quad\overline{\gamma}  \big|h_{d,b_{n,c}}^c(n)+\sum\limits_{u=1}^{2^{J_n}} {h}_{b_{n,c},c,u}^{J_n}(n) \overline{\phi}_{u,c}^*
 \big|^2 \big)\Big).
 \vspace{-4mm}
\end{align}
\vspace{-1mm}
And the RIS-aided transmission time is $(T_c\!-\!\tau_{\rm{CE}}(J_n))$.
\vspace{-1mm}
\section{Optimal RIS-aided RA Strategy}\label{s:probem_solution}
\vspace{-1mm}
\subsection{Decision-Making Problem Formulation}
In this section, we cast the random access problem during one time frame by following the proposed multiple phased RA scheme as a sequential decision optimization problem.

For RR phase $n\in \mathbb{N}$, let $(\psi_n,\varphi_n)$ and $\eta_n$ denote the BS's two layers decisions after $n$th RR phase, as described in Sec. \ref{sub:proposed_scheme}.
Specifically, for 1st layer decision, $\psi_n=0,{\overline{J}}+1$ or $1,\ldots,\overline{J}$ refer to options 1), 2) and subarray grouping level $J_n\in[1,\overline{J}]$ of option 3), correspondingly; $\varphi_n\in \{0,1,...,K\}^C$ refers to users probing vector $\mathbf{b}_n$.
Moreover, for 2nd layer decision, by following option 3) with decision $\eta_n$ and $\varphi_n$, $\eta_n=0$ or $1$ refers to RIS-aided data transmission or next RR phase, respectively.

Correspondingly, we denote $Y_n$ and $T_n$ as the transmitted traffic volume during the DT phase and 
total duration after $n$th RR phase, respectively.
The traffic volume is expressed as 
\vspace{-5mm}
\begin{align*}
    Y_n=&\mathbb{I}[\psi_n=0]\cdot R_{d,n}^*T_c+\\&\mathbb{I}[1\le \psi_n \le \overline{J}]\cdot\mathbb{I}[\eta_n=0]\cdot R_{r,n}(\varphi_n) (T_c-\tau_{\rm CE}(\psi_n)).
\end{align*}
The total duration is expressed as 
\begin{equation*}
\setlength\abovedisplayskip{1pt}
\setlength\belowdisplayskip{7pt}
T_n=\sum_{l=1}^n \tau_{\rm RA} +\sum_{l=1}^{n-1}\mathbb{I}[1\le \psi_l \le \overline{J}]\cdot \tau_{\rm CE}(\psi_l)+T_c.
\vspace{-2mm}
\end{equation*} 
Moreover, as the RR phase for data transmission is random, decision sequence $(\psi_n,\varphi_n,\eta_n)$ $n\in\mathbb{N}$ determines the stopping time by
\vspace{-2mm}
\begin{equation*}
\setlength\belowdisplayskip{1.7pt}
N=\inf\{n>0: \psi_n=0,~\text{or}~1\le \psi_n \le {\overline{J}}~\text{and}~\eta_n=0\}.
\end{equation*} 
As realizations of observed information $\{{\mathcal I}_n,\mathbf{h}_{d}(n),$ $\mathbf{f}_{k,c}(n),\mathbf{g}_{k,c}(n)\}$ and random, the decision sequence $(\psi_n,\varphi_n,\eta_n)$ and the stopping time $N$ are both random. 
The average system throughput $\lambda_N$ 
is calculated as
\vspace{-2mm}
\begin{equation}
\setlength\belowdisplayskip{1.9pt}
	\lambda_{N}=\frac{\mathbb{E}[\text{Data volume in one frame}]}{\mathbb{E}[\text{Duration of one frame}]}=
 \frac{\mathbb{E}[Y_{N}]}{\mathbb{E}[T_{N}]}
\end{equation}
Therefore, the maximal average system throughput and corresponding optimal strategy can be expressed as\vspace{-2mm}
\begin{equation}
\setlength\belowdisplayskip{1.9pt}
    \lambda^*=\sup_{N> 0} \frac{\mathbb{E}[Y_{N}]}{\mathbb{E}[T_{N}]}, 
    (\psi_n^*,\varphi_n^*,\eta_n^*,N^*)=\arg\sup\limits_{N>0} \frac{\mathbb{E}[Y_{N}]}{\mathbb{E}[T_{N}]}
\end{equation}
For notation simplicity, in the sequel we use stopping time $N$ to denote sequential decision strategy $(\psi_n,\varphi_n,\eta_n), n\in{\mathbb{N}}$ and optimal RA strategy, while $N^*$ denotes the optimal strategy.
\vspace{-2mm}
\subsection{Preliminary of Sequential Plan Decision Theory}\label{sec_optSPD}
The optimal sequential plan decision (SPD) theory uses dynamically planned sequential samples based on accumulated observation information~\cite{Zhou2024_globecom},\cite{Stadje1997}. 
The classic sequential decision optimization problem assumes that \(\mathbf{X}_n, \mathbf{X}_n^\dag, n \in \mathbb{N}\), $n\ge 1$ are i.i.d. sequence of random triplets, and $Z_n, n\ge 1$ is the reward when stopping at time $n$. For each $\mathbf{X}_n^\dag$, we denote $\mathbf{X}_n^\dag=\{\mathbf{X}_{n,j}^\dag\}_{j=1,...,J}$, with $\mathbf{X}_{n,j}^\dag=\{X_{n,j}^\dag(1),...,X_{n,j}^\dag(A)\}$.
The successive observation of sequence $\mathbf{X}_1,\mathbf{X}_1^\dag,\mathbf{X}_2,\mathbf{X}_2^\dag,...$ with option of skipping any $\mathbf{X}_n^\dag$ of or observing a portion matters the
decision process. Proceeding from time $n$ to $(n\!+\!1)$ costs $c_0$, and obtaining additional information $\mathbf{X}_{n,j}^\dag$ at time $n$ costs $d_0(j)$.

An SPD strategy is a triple of functions $\psi_n: \mathbb{R}\to \{0,1,...,\overline{J}+1\}$, $\varphi_n: \mathbb{R}^C\to \{0,1,...,K\}^C$, and $\eta_n: \mathbb{R}\to \{0,1\}$.
At each time \(n\), if reward $Z_{n,1}(\mathbf{x}_1,\mathbf{x}_1^\dag,...,\mathbf{x}_n)=z_1$ is observed, $\psi_n(z_1)=0$ or $1$ means the  acceptance or rejection  for stopping at time $n$, and $\psi_n(z_1)>1$ means additional
observation of $\mathbf{X}_n^\dag$ is called for. In the case $\psi_n(z_1)>1$, $\psi_n(z_1)=j$ and $\varphi_n(z_1)=\mathcal{A}$ mean a subset of $\mathbf{X}_{n,j}^\dag$ is observed.
Then, in the case $\psi_n(z_1)=j$ and $\phi_n(z_1)={\mathcal A}$, 
if reward $Z_{n,2}(\mathbf{x}_1,\mathbf{x}_1^\dag,...,\mathbf{x}_n,\mathbf{x}_n^\dag)=z_2$ is observed, $\phi_n(z_2)=0$ or $1$ means the  acceptance or rejection for stopping at time $n$.
By following the strategy $(\psi_n,\varphi_n,\eta_n)$, the reward is obtained by stopping at time $N$, where  $N=\inf\{n>0: \psi_n=0,~\text{or}~1\le \psi_n \le {\overline{J}}~\text{and}~\eta_n=0\}$.
Thus, the expected reward is 
\vspace{-2mm}
\begin{equation*}\label{equ:general_goal}
\setlength\abovedisplayskip{1pt}
\setlength\belowdisplayskip{1pt}
     \mathbb{E}[Z_N]=\mathbb{E}\big[Z_{N,1}\mathbb{I}[\psi_N\!=\!0]+Z_{N,2}\mathbb{I}[1\le \psi_N \le \overline{J}]\mathbb{I}[\eta_N\!=\!0]\big].
\end{equation*}
The goal is to achieve the maximal reward $U_0=\sup\limits_{N>0}\mathbb{E}[Z_N]$.

Based on SPD theory, an expectation-threshold based strategy is derived achieving $\sup_{N}\mathbb{E}[Z_N]$ in the following lemma.
 
\begin{lemma}\label{th:optimal_rule2}
The optimal decision strategy $N_{T}^*$ achieving $\sup_{N}\mathbb{E}[Z_N]$ is that: starting from time $n=0$, 
it is optimal 
\begin{enumerate}
    \item to stop at time $n$ with $\psi_n^*=0$ if $Z_{n,1}\ge \max_{j,\mathcal{A}}\mathbb{E}\big[\max\{Z_{n,2},U_0\}|\mathbf{X}_{n}\big]$ 
    and $Z_{n,1}\ge U_0 $;
    \item to reject stopping if $Z_{n,1}<\max_{j,\mathcal{A}}\mathbb{E}\big[\max\{Z_{n,2},U_0\}|\mathbf{X}_{n}\big]$ and $Z_{n,1}< U_0 $;
    \item to observe additional information $\mathbf{X}_{n,\psi_n^*}^\dag(\varphi_n^*)$ with $(\psi_n^*,\varphi_n^*)$ such that $\mathbb{E}\big[\max\{Z_{n,2}(\psi_n^*,\varphi_n^*),U_0\}|\mathbf{X}_{n}\big]-d_0(\psi_n^*)=\max_{j,\mathcal{A}}\big\{\mathbb{E}\big[\max\{Z_{n,2}(j,\mathcal{A}),U_0\}|\mathbf{X}_{n}\big]-d_0(j)\big\}$, and then to stop at time $n$ with $\eta_n^*=0$ if $Z_{n,2}\ge U_0$, and otherwise to reject with $\eta_n^*=1$.
\end{enumerate} 
Moreover, the maximal reward $U_0$ can be solved by 
\vspace{-2mm}
\begin{equation*}
U_0=\mathbb{E}\big[\max\{Z_{n,1},U_0,\max\limits_{j,A}\mathbb{E}\big[\max\{Z_{n,2},U_0\}|\mathbf{X}_{n}\big]\}\big]-c_0.
\end{equation*}
\vspace{-5mm}
\end{lemma}
\begin{IEEEproof}
It can be proved similar to Theorem 1 in \cite{Stadje1997}.
\end{IEEEproof}

\vspace{-2mm}
\subsection{Optimal RIS-aided RA Strategy}
\vspace{-1mm}
In this section, we use the optimal decision strategy $N_{T}^*$ to derive optimal RRIS-aided RA strategy $N^*$ maximizing the average throughput $\frac{\mathbb{E}[Y_N]}{\mathbb{E}[T_N]}$.

To bridge the equivalence between the reward and fractional maximization problems, we have the results as follows. 
\begin{lemma}\label{l:equ_problem_trans}
An optimal strategy denoted as $N^*(\lambda^*)$ maximizing the average cost price reward $\mathbb{E}[Y_{N}-\lambda^* T_{N}]$ such that $\sup_{N}\mathbb{E}[Y_{N}-\lambda^* T_{N}]=0$ is an optimal strategy $N^*$ maximizing average throughput $\sup_{N>0} \frac{\mathbb{E}[Y_{N}]}{\mathbb{E}[T_{N}]}$. Moreover, 
 $\lambda^*$ uniquely exists such that $\sup_{N}\mathbb{E}[Y_{N}-\lambda^* T_{N}]=0$.
\end{lemma}

Based on results of Lemma \ref{th:optimal_rule2} and Lemma \ref{l:equ_problem_trans}, we derive the optimal RA strategy as follows.
We define threshold function\footnote{Notably, as the i.i.d. statistical property of observed CSI,
for $\forall n\ge 1$, right-hand-side of 
function is equal and thus time index $(n)$ is omitted.} 
\vspace{-4mm}
\begin{equation*}\label{equ:thres_func}
\setlength\abovedisplayskip{1pt}
\setlength\belowdisplayskip{0pt}
        {\Theta}_{(\mathbf{b},J)}(\lambda,\mathbf{h}_d)
	{=}\mathbb{E}\big[\max\big\{
	T_{r,J} R_r(\mathbf{b})	-\lambda T_c,-\lambda \tau_{\rm{CE}}(J)
	\big\}\big|\mathbf{h}_{d}\big].
\end{equation*}
where $T_{r,J} = T_c-\tau_{\rm{CE}}(J)$. It represents the expected reward corresponding to option 3) by letting users estimates the RIS by probing vector $\mathbf{b}$ under subarray grouping level $J$, when user-BS link channel gains $\mathbf{h}_d$ have been observed.
\begin{theory}\label{th:REB_rule1}
The optimal RA strategy $N^*$ achieving $\sup_{N>0}\frac{\mathbb{E}[Y_N]}{\mathbb{E}[T_N]}$ is as that:
after $n$th RR phase, BS obtains $\{\mathcal{I}_n,\mathbf{h}_{d}(n)\}$ and makes 1st layer decision:
\begin{enumerate}
    \item if $(R_{d,n}^*-\lambda^*) T_c\ge \max\big\{\max\limits_{\mathbf{b},J} {\Theta}_{(\mathbf{b},J)}({\lambda^*},\mathbf{h}_{d}(n)),0\big\}$, select option 1) with $\psi_n^*=0$;
    \item if $\max\big\{(R_{d,n}^*-\lambda^*) T_c,\max\limits_{\mathbf{b},J} \Theta_{(\mathbf{b},J)}\big(\lambda^*,\mathbf{h}_{d}(n)\big)\big\}<0$, select option 2) with $\psi_n^*=\overline{J}+1$;
    \item else, select option 3) with $\psi_n^*=J_n^*$, $\varphi_n^*=\mathbf{b}_n^*$, where 
    \begin{equation*}
    \vspace{-1mm}
    \setlength\abovedisplayskip{1pt}
        (J_n^*,\mathbf{b}_n^*)=\arg\max\limits_{\mathbf{b}_n,J_n} \Theta_{(\mathbf{b}_n,J_n)}\big({\lambda^*},\mathbf{h}_{d}(n)\big).
         \vspace{-1mm}
    \end{equation*}
    After additional observations following decisions $(\psi_n^*,\varphi_n^*)$, if $R_{r,n}(\varphi_n^*) \ge \lambda^* $, select option 1) with $\eta_n^*=0$, otherwise, select option 2) with $\eta_n^*=1$.
\end{enumerate}
Moreover, the maximal throughput $\lambda^*$  uniquely satisfies 
\vspace{-3mm}
\begin{equation}\label{equ:ergodic_cap2}
\vspace{-4mm}
\mathbb{E}\big[\!\max\big\{(R_{d}^*\!-\!\lambda^*)T_c,0,
	\max_{\mathbf{b},J}
	 \Theta_{(\mathbf{b},J)}({\lambda^*},\mathbf{h}_d)\big\}\big] \!=\!\lambda^*\tau_{\rm{RA}}.
\end{equation}
\end{theory}\vspace{1mm}
\begin{IEEEproof}
It can be proved similar to Theorem 2 in \cite{Zhou2024_tvt} and Theorem 1 in \cite{Zhou2025_tmc}.
\end{IEEEproof}

 \vspace{-2mm}
\section{RIS-aided RA Algorithm}
 \vspace{-2mm}
Based on the proposed optimal strategy $N^*$, we develop the RA algorithm for system operation.
\begin{algorithm}
	\caption{Iterative algorithm for $\lambda^*$}\label{Algorithm1}
	\SetKwInOut{Input}{Input}\SetKwInOut{Output}{Output}
	\Input{$\lambda_0=1$, $\Delta=1$, $l=1$, $\epsilon$}
	\Output{$\lambda^*$}
	\While{$\Delta \ge \epsilon$}{
    \For{$I=1$ \KwTo $S$}{
    $\mathbf{h}_I \leftarrow [\mathbf{h}_{d,1}, \dots, \mathbf{h}_{d,I}]$ \\
    $\Lambda_{I} \leftarrow  \mathbb{E}_{\mathbf{h}_I} \big[
     \max \big\{ T_c(R_d^*(\mathbf{h}_I)-\lambda_l),
     0, \max_{\mathbf{b}, J} \Theta_{(\mathbf{b}, J)}(\lambda_l, \mathbf{h}_I)
     \big\}
    \big]$ 
    }
    Update $\lambda_{l+1} \leftarrow \lambda_l + \alpha \left(\sum_{I=1}^S p_I \Lambda_{I} - \lambda_{l} \tau_{\rm{RA}} \right)$, \\
			where $p_I = \binom{S}{I} \frac{K^I (S-1)^{I(K-1)} \left(S^K - K(S-1)^{K-1}\right)^{S-I}}{S^{KS}}$ \\
			and step-size $\alpha$ satisfies $\epsilon \le \alpha \le \frac{2 - \epsilon}{\tau_{\rm{RA}} + T_c}$. \\
			$l \leftarrow l + 1$
    }
    	$\lambda^* \leftarrow \lambda_{l}$ 
\end{algorithm}

First, based on the equation \eqref{equ:ergodic_cap2}, we design an iterative algorithm as specified in Algorithm~\ref{Algorithm1} to calculate throughput $\lambda^*$ in an offline manner, where $\epsilon$ denotes numerical accuracy.
The convergence of the iterated sequence $\{\lambda_l\}, l=1,...,\infty$ is guaranteed by the following theorem. 
\begin{theory}
The generated sequence $\{\lambda_l\}, l=1,...,\infty$ by Algorithm \ref{Algorithm1} converges to the maximal throughput $\lambda^*$.
\end{theory}
\begin{IEEEproof}
It is proved by Proposition 1.2.3 in \cite{Berts1999}.
\end{IEEEproof}
 
Then, using throughput $\lambda^*$ from Algorithm~\ref{Algorithm1} as the input, we develop the online RA algorithm.
The following findings can be used to decrease complexity for 1st layer decision.
\begin{lemma}\label{l:complx_reduce}
   For the 2nd layer decision after $n$th RR phase, the optimal users probing vector $\varphi_n^*$ is equal to the optimal scheduling vector $\mathbf{a}_n^*$ for the 1st layer.
\end{lemma}
\begin{IEEEproof}
It can be proved by contradiction using increasing monotonicity of $\Theta_{(\mathbf{b},J)}(\lambda^*,\mathbf{h}_d)$ over $h_{d,k}^c$. 
\end{IEEEproof}





Based on Lemma \ref{l:complx_reduce} and Theorem 1, we design an RIS-aided RA algorithm in Algorithm \ref{Algorithm2} by following the proposed RA scheme in Sec. \ref{sub:proposed_scheme}.
The algorithm has a pure-threshold structure, which only requires system statistics and can be obtained using pre-computed values stored in a look-up table.
 \vspace{-3mm}
\begin{algorithm}
	\caption{RIS-aided RA algorithm}\label{Algorithm2}
	\SetKwInOut{Input}{Input}\SetKwInOut{Output}{Output}
	\Input  { ${\lambda^*}$}
	 \vspace{-1mm}
	\Repeat{all data transmissions finish}{
		after one RR phase, BS obtains granted users $\mathcal{I}$ and link CSI $\mathbf{h}_{d}$.

        find optimal scheduling vector $\mathbf{a}^*$ by Alg. 1
        
        calculate user-BS transmission rate $R_{d}^*$ by (1)


        calculate $\overline{\Theta}\leftarrow\max\limits_{J} \Theta_{(\mathbf{a}^*,J)}({\lambda^*},\mathbf{h}_d)$        
  
			\uIf{$R_{d}^*\ge \lambda^* ~\&~ R_{d}^*\ge \lambda^*+\overline{\Theta}/T_c$ }{
				schedule users by vector $\mathbf{a}^*$ to transmit in user-BS link with rate $R_{d}^*$ and duration $T_c$.}
			\uElseIf{ $R_{d} < \lambda^* ~\&~R_{d}<\lambda^*+\overline{\Theta}/T_c$}{
				\vspace{-1mm}
				skip to new RR phase.}
			\Else{ \vspace{-1mm}
   $J^*\leftarrow \arg\max\limits_{J} \Theta_{(\mathbf{a}^*,J)}({\lambda^*},\mathbf{h}_d)$,~$B\leftarrow2^{\overline{J}-J^*}$
				
               let users to probe RIS by vector $\mathbf{a}^*$ under subarray grouping level $J^*$ in a GRP phase

    \For{$c=1:C$}
				{calculate optimal beamforming vector $\overline{\mathbf{\Phi}}_c^*=[\overline{\phi}_{1,c}^*,\ldots,\overline{\phi}_{2^{J^*},c}^*]^T$ 
    }
               calculate RIS-aided rate $R_{r}^*$ by (3)
				
				\uIf{$R_{r}^* \ge {\lambda^*}$}{
				schedule users by $\mathbf{a}^*$ to transmit aided by the RIS with rate $R_{r}^*$ and beamforming vector $\overline{{\Phi}}_c^*\otimes\mathbf{1}_{B\times1}$
				during $(T_c-\tau_{\rm{CE}}(J^*))$.}
				\Else{ 
				skip to new RR phase.
    \vspace{-1mm}}
			}
	}
\vspace{-1mm}
\end{algorithm}
 \vspace{-3mm}

\begin{figure*}[t]
	\centering
	\subfloat[Average throughput vs $P_t$.]{
		\label{fig:comparison1}
		\includegraphics[width=0.33\textwidth]{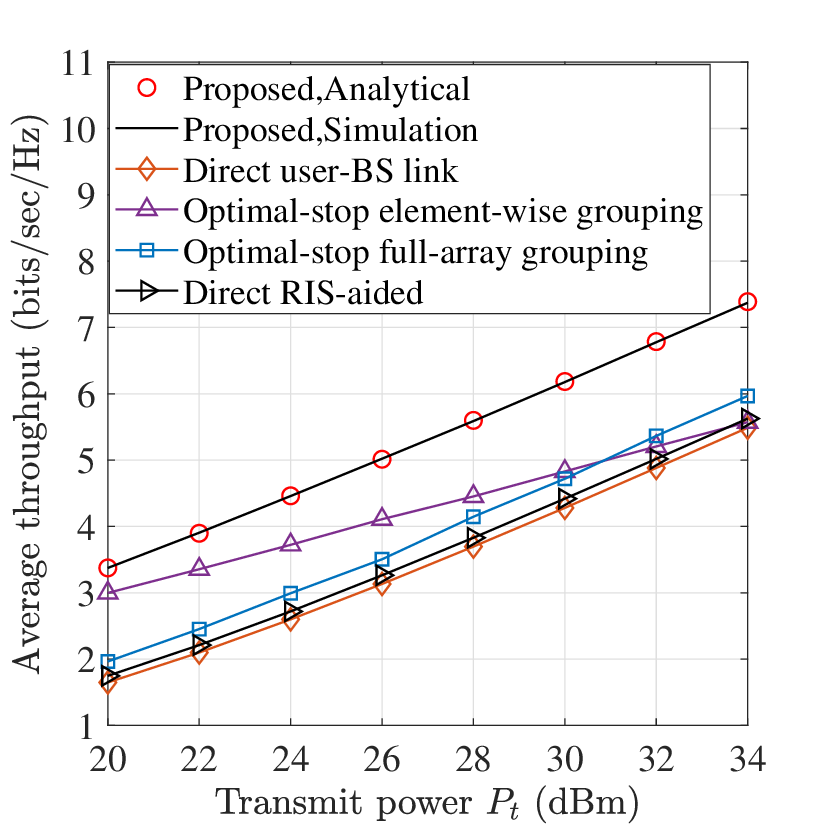}} 
	\subfloat[Average throughput vs $T_c$. ]{
		\label{fig:comparison2}
		\includegraphics[width=0.33\textwidth]{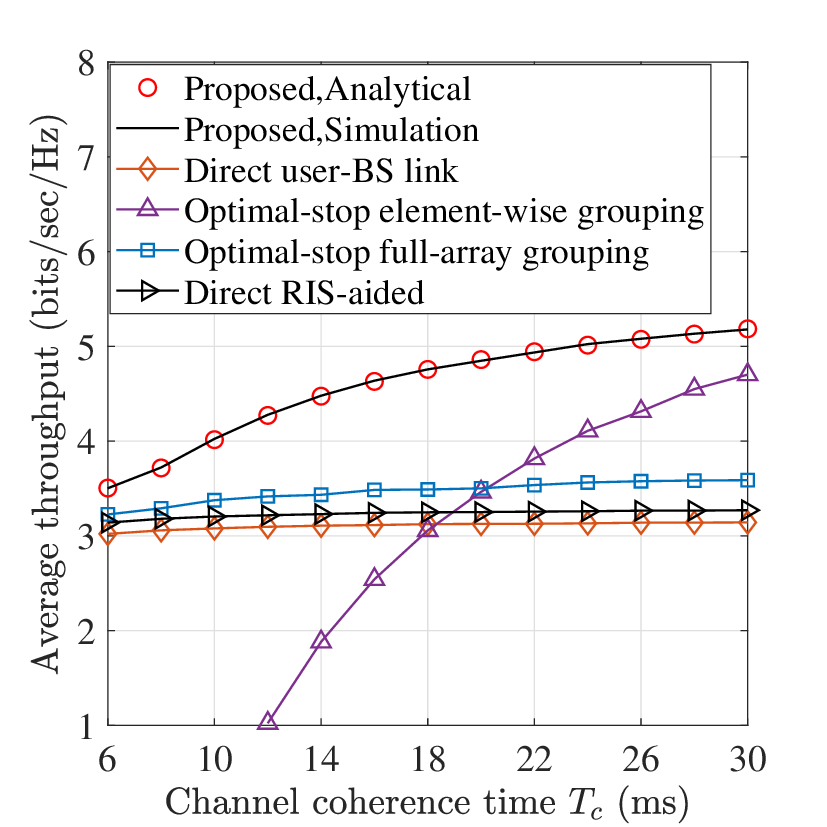}} 
  \subfloat[Average throughput vs $M$.]{
		\label{fig:comparison3}
		\includegraphics[width=0.33\textwidth]{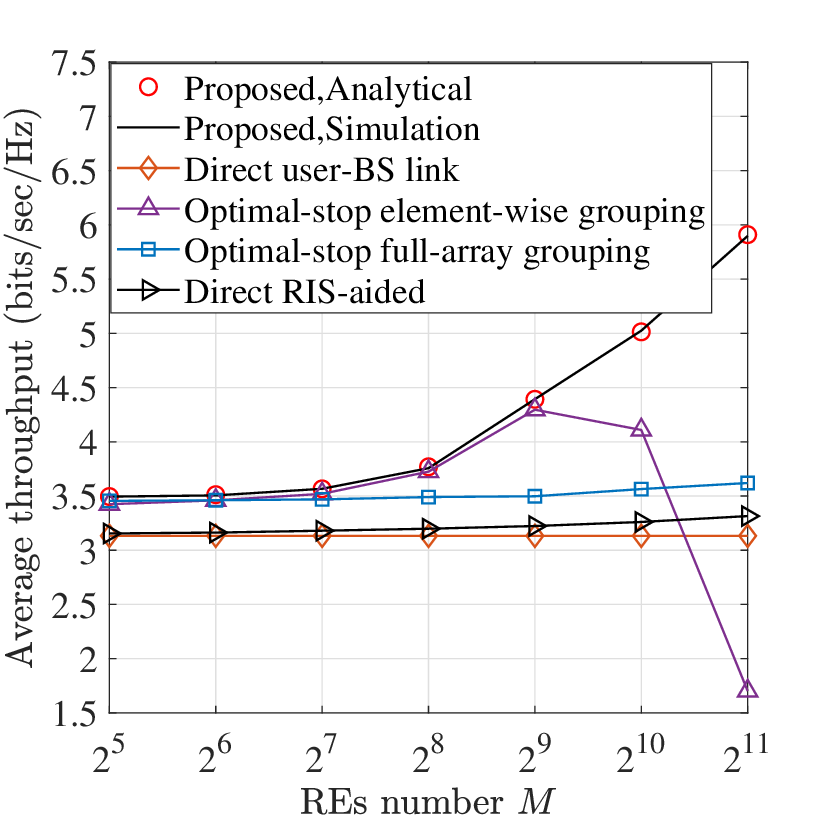}}
         \vspace{-1mm}
     \caption{Validation and performance comparison with alternative strategies. 
     }
	\label{fig:comparison}\vspace{-5mm}
\end{figure*}

 \vspace{-0.3mm}
\section{Numerical Results}\label{s:num}
 \vspace{-0.3mm}
We consider a cellular network with $K=50$ users, preamble size $S=30$, and $C=8$ sub-channels. The BS is located at $(0,100,0)$, the RIS at $(200,0,20)$, and users are centered at $(1000,100,0)$ in 3D Cartesian coordinates. The system operates at $f_c = 2$ GHz with a reference link loss of $\beta_0 = -30$\,dB at $1$\,m, and noise power $N_0 = -100$\,dBm. 
The RR phase duration is $\tau_{\rm{RA}} = 0.3$\,ms, and the pilot duration in the GRP phase for one RE is $\tau_s = 10\,\mu$s. To validate the theoretical results, Fig.~3 compares the analytical results from Algorithm~$2$ with the simulation results from Algorithm~$3$ across varying parameters. 
The close match confirms the accuracy of our proposed strategy.

We compare the performance of the proposed strategy with four alternative strategies from the literature: 
i) \textit{Direct user-BS link} strategy: After each RR phase, the BS schedules at most $C$ users and lets them transmit data using OFDMA without RIS assistance~\cite{Althumali2020}.
ii) \textit{Direct RIS-aided} strategy: After each RR phase, the BS schedules at most $C$ users and divides the RIS into equal-size subarrays. Users probe the RIS under the full-array grouping and transmit data using RIS-aided orthonormal sub-channels~\cite{CaoMar2021}.
iii) \textit{Optimal-stop element-wise grouping} strategy: After each RR phase, the BS obtains both user-BS link and RIS CSI acquired by probing the RIS element-wisely under the maximal
subarray grouping level $\overline{J}$. An optimal stopping strategy is used to decide when to schedule users for RIS-aided transmission under 
optimal beamforming, combining strategies from \cite{Kundu2022}, \cite{Zhou2023_globecom} and \cite{Wei2020acm}.
iv) \textit{Optimal-stop full-array grouping} strategy: After each RR phase, the BS obtains both the direct link CSI and RIS CSI acquired by probing under the minimal subarray grouping level with subarray of $2^{\overline{J}-1}$ REs. An optimal stopping strategy is then applied for RIS-aided transmission, combining strategies from \cite{Kundu2022}, \cite{Zhou2024_icc} and \cite{Wei2020acm}.
In all strategies, optimal users are selected across sub-channels following Algorithm~$1$ from detected users. 

Fig.~\ref{fig:comparison1} compares the average throughput of the proposed strategy with alternatives across transmit power $P_t$ at $T_c=24$ ms and $M=2^{10}$. As $P_t$ increases from 20 to 34\,dBm, the proposed strategy outperforms all alternatives, achieving a 12.6\% throughput gain over the optimal-stop element-wise grouping strategy (the best among alternatives) and a 31.3\% gain over the no-wait-direct strategy (the worst).
Moreover, Fig.~\ref{fig:comparison2} shows the average throughput versus channel coherence time $T_c$ at $P_t = 26$ dBm and $M = 2^{10}$. The proposed strategy significantly outperforms all alternatives. As $T_c$ increases from 6 ms to 30 ms, the performance advantage over the optimal-stop element-wise grouping strategy decreases, while advantages over other strategies increase. This is due to RIS-aided gains dominating the RIS probing overhead, with higher subarray grouping levels leading to higher throughput. The advantage over the optimal-stop element-wise grouping strategy is due to the flexibility of the proposed strategy in dynamically adapting subarrays.
In addition, Fig.~\ref{fig:comparison3} shows the throughput versus RIS size $M$ at $T_c \!=\! 24$\,ms and $P_t \!=\! 26$\,dBm. The proposed strategy outperforms all alternatives, with performance advantage increasing as $M$ grows. This is due to a better balance between transmission gain and probing overhead. The optimal-stop element-wise grouping strategy performs similarly to the proposed strategy when $M < 2^9$, but its performance degrades as $M$ increases due to heavy overhead from probing all REs.

  \vspace{-2mm}
\section{Conclusion}\label{s:con}
 \vspace{-2mm}	
This study presents a novel opportunistic massive random access (RA) framework for distributed users in a subarray grouping RIS-aided cellular network. By applying sequential decision optimization theory, we developed an optimal expectation-threshold-based strategy, implemented as a multi-phase RA scheme. This approach maximizes system throughput by exploiting the time-invariant properties of observation processes, enhanced RIS performance through subarray grouping, and leveraging multi-user and time diversity. We proposed an offline iterative algorithm for throughput calculation and an online RA algorithm for efficient execution, ensuring practical applicability. The proposed framework significantly improves multi-user transmission throughput, offering enhanced performance with manageable overhead. 


\end{document}